# Homogenous and Heterogenous Parallel Clustering: An Overview

Ahmed Ibrahim
Computer Science Department
Boston University, MA, USA
London, Canada
aibrahim@bu.edu

Rokaya Hassanien
Electrical, Computer, and Biomedical Engineering
Ryerson University
Toronto, Canada
rhassanien@ryerson.ca

*Abstract*—Recent advances in computer architecture and networking opened the opportunity for parallelizing the clustering algorithms. This divide-and-conquer strategy often results in better results to centralized clustering with a much-improved time performance. This paper reviews key parallel clustering and provides insight into their strategy. The review brings together disparate attempts in parallel clustering to provide a comprehensive account of advances in this emerging field.

## 1. Introduction

Many businesses have adopted various machine learning approaches [1]-[15], with varying definitions of clusters, classes, boundaries, constraints, and similarity measurements. Due to the massive volume of data in diverse applications such as data mining and bioinformatics, the issue of scalability and dealing with large and high dimensional datasets is particularly significant. Clustering such big and high-dimensional datasets in unsupervised machine learning is a current difficulty, making typical centralized clustering algorithms ineffective. Clustering techniques must be parallelized by separating computations so that each node can do a piece of the clustering operation in parallel with the other processing nodes. As a result, faster results are obtained than with a centralized node system. On both the local and global levels, parallel clustering is carried out. On a local level, nodes cluster autonomously, and local models (i.e., representatives) are generated at each node. After the local clustering is completed, a global model is created. Parallelization of clustering algorithms can be categorized into main categories, homogeneous or heterogeneous parallelization depending on the types of clustering algorithms used. An overview of several subjects that fall under the umbrella of parallel data clustering is covered in this study to provide a thorough account of advances in these new disciplines. The following is a breakdown of the paper's structure: Section 2 discusses several clustering parallelization algorithms. Section 3 discusses future smart systems using parallelization strategies. Finally, in section 4, we wrap up the paper with a summary.

# 2 Parallel Data Clustering

Efficient parallel and large-scale algorithms [16]-[31] minimize the calculation time while clustering big data sets. Splitting the clustering task among P nodes is required for parallel data clustering. The computing process must be balanced for each node in clustering parallelization; the balancing strategy reduces communication costs during the clustering phase.

## 2.1 Parallel Homogenous Clustering Algorithms

This section focuses on homogenous parallel clustering algorithms. Each node invokes the same clustering strategy, and homogenous local prototypes are generated at each node.

### 2.1.1 Parallel *k*-Means (PKM) Algorithm

The PKM [24][25] is an iterative parallel version of the KM using MPR reduction routines [32].

---
**Algorithm: Parallel *k*-means (PKM)**
**Inputs:** The dataset X, # of clusters *k*, # of nodes P
**Outputs**: *k* clusters
- Split the dataset X on the P nodes.
- $N_0$ randomly chooses initial centroids and replicates them to other P-1 nodes.
  **Repeat**
   *Stage1*: local nodes $N_p$ calculates the distance of the local vectors to the *k* centres.
   *Stage2*: Local vectors are mapped to the closest centroid, the local *objective function* is evaluated
   *Stage3*: The local representatives are reduced to produce the global representatives using MPI [32]
  **Until Convergence**
---

**Fig. 1**. The PKM Algorithm

### 2.1.2 Parallel Fuzzy *c*-Means (PFCM) Algorithm

The PFCM [26] presented in Fig.2 distributes the fuzzy membership function *on P* nodes with a local model containing only the local memberships. The global reduction of local centroids produces global centroids with calls to MPI reduction routines [32]. This step is repeated until convergence is determined by minimizing the distance to global centroids.

---
**Algorithm: Parallel Fuzzy c-means (PFCM)**
**Inputs**: The dataset X, the weighting exponent *m*, # of clusters *k* and # of nodes P
**Outputs**: *k* clusters
Distribute X on the P nodes such that each node calculates the initial local membership matrix
  **Repeat**
   *Stage1*: Node $N_p$ calculates centroids for the *k* clusters with global reductions with 2 MPI calls.
   *Stage2*: Node $N_p$ computes the local membership of each local vector to the *k* clusters using the updated global *k* centroids and assigns local points to new clusters
   *Stage3*: Global reduction is performed to produce global *J* value
  **Until Convergence**
---

**Fig. 2**. The PFCM Algorithm

### 2.1.3 Parallel *k*-Windows (PK-windows) Algorithm

The *k*-windows algorithm uses the concept of windows to determine clusters with the Orthogonal Range Search (ORS) algorithm described in Fig.3. In the PK-windows algorithm [21], the range search procedure requires the most computational effort; thus, a parallel algorithmic scheme that uses a Multi-Dimensional Binary Tree [33] with a Server-Slave model for a range search [34] is presented in Fig.4.

**Algorithm: Orthogonal Range Search (ORS)**
**Inputs:** The tree *TR*, the i$^{th}$ coordinate and the *d*-range query *Q*
**Outputs:** SS
*SS*= { }
*Stage1*: $x_r$ is the root of *TR*
*Stage2*: Examine the tree *TR*, if $x_r \in Q$, then Add $x_r$ to *SS*, else recursively explore the left and right trees of $x_r$ wrt to next coordinate *i*+1

**Fig.3**. The ORS

**Algorithm: Parallel *k*-windows**
**Inputs:** The dataset X, # of windows *l*, and the *d*-ranges area *a*.
**Outputs:** *k* clusters
The *l* centroids are chosen with initial *d*-ranges windows $W_i$, *i*=1,..,*l* centred on these initial centroids with area *a*.
*Phase1*: **Repeat**
    *Stage 1*: **Repeat**
        -The master $N_i$; sends a "*start new sub-search*" message to the ideal slave $N_j$; $i \neq j$
        -At the slave node, if a new sub-search is necessary,
            The slave node $N_j$ sends a "*necessary new sub-search*" message to the master node $N_i$ and spawns to another ideal node, otherwise
            Return the set of vectors that lie within the given *d*-range query Q.
    **Until no new sub-search messages are desired**
    *Stage 2*: New centroids are calculated as new *d*-ranges
**Until each window has no significant increment of vectors**
*Phase2*, *Phase3* and *Phase4* are typical in the classical *k*-windows algorithm [21].

**Fig.4** The Parallel *k*-windows Algorithm

### 2.1.4 Distributed Clustering using Principal Component Analysis (DCPCA) Algorithm

Figure 5 depicts the Collective Principal Component Analysis (CPCA) process. The computations for PCA could be done locally in a parallel environment, reducing the quantity of data transfer and processing at one central node. The distributed clustering approach [35] works with a provided centralized algorithm B module and respects the user's choice of any local clustering algorithm. Figure 6 depicts a distributed CPCA-based system.

**Algorithm: Collective Principal Component Analysis (CPCA)**
**Inputs:** The distributed vertical partitioned local data $X_p$
**Outputs:** Global Principal Components of the original data X
*Stage1*: Node $N_p$ performs a PCA algorithm locally and selects dominant eigenvectors.
*Stage2*: Each node $N_p$ sends projected data with the eigenvectors to the facilitator
*Stage3*: The facilitator combines the data received from all the nodes and performs PCA on the global set, identifies the dominant eigenvectors, and transforms them to the original space.

**Fig.5**. The CPCA Algorithm

**Algorithm: Distributed Clustering CPCA-based**
**Inputs:** The distributed datasets $X_p$; $p=0,1,...,P-1$
**Outputs:** $k$ clusters
*Stage1*: Node $N_p$ performs PCA locally and projects the local data on the local PCs
*Stage2*: Node $N_p$ applies the clustering algorithm B.
*Stage3*: Each cluster's representative points are selected at every node $N_p$. Let $I_p$ is the set of indices at node $N_p$, i.e., the chosen representative points.
*Stage4*: All nodes communicate and send the data rows to the facilitator node.
*Stage5*: The facilitator performs the global PCA and transmits the global PCs to all nodes.
*Stage6*: Node $N_p$ maps the local data to the global PCs and performs B clustering.
*Stage7*: Node $N_p$ communicates and sends a sketch of the local clusters to the facilitator node.
*Stage8*: The facilitator combines the different local descriptions (models) obtained from the local nodes to construct global clusters.

**Fig.6**. The Distributed Clustering CPCA-based Algorithm

### 2.1.5 Distributed Density-Based Clustering (DDBC) Algorithm

In the DDBC algorithm [36], the local model is created using two local models, $REP_{k\text{-means}}$ and $REP_{scor}$ [36][37], using the notion of specified core points. The $REP_{k\text{-means}}$ based on local specific core points at node $N_p$ is acquired using the $k$-means as shown in Fig.7. Each local cluster Si in REPscor is characterized by a full set of specific core points ScorSi.

**Algorithm: DBSCAN Clustering using *k*-means**
**Inputs:** local cluster $S_i$, $i=1,2,...,k$, using centralized DBSCAN at local nodes $N_p$
**Outputs** $REP_{k\text{-means}}$
*Stage1*: Apply the local clustering $S_i$ runs on the $N_p$, re-clustered using $k$-means, the set of $|Scor_{Si}|$ centroids $c_{i1},c_{i2},...,c_{i|ScorS|}$ is shaped
*Stage2*: centroid $c_{ij}$; $j=1,2,...,|Scor_{Si}|$ is allocated with a ε-range value to indicate the area $c_{ij}$.
*Stage3*: The k clusters will have a local model as:
$$LocalModel_p = \bigcup_{i=1,...,k} \bigcup_{j=1,...,|Scor_{Si}|} \{(c_{ij},\varepsilon_{c_{ij}})\}$$

**Fig. 7**. Local DBSCAN using *k*-means

Each local model comprises *k* local clusters with matching representatives, and the facilitator node creates a global model using local models. Every local representative develops its own cluster in the DBSC algorithm. Figure 8 depicts the DDBC algorithm.

**Algorithm: Distributed Density-Based Clustering**
**Inputs:** local data $X_p$, , $Eps_{local}$, and $MinPts_{Local}$
**Outputs:** $k$ clusters
*Stage1*: Node $N_p$ execute the DBSCAN using $Eps_{local}$ and $MinPts_{Local}$ and generates local models
*Stage2*: $N_p$ sends the local model models to the facilitator node.
*Stage3*: facilitator executes a global DBSCAN with global $MinPts_{global}$ and $Eps_{global}$
*Stage4*: facilitator sends the global model to nodes $N_p$.
*Stage5*: Node $N_p$ re-assign local points, the two clusters (i.e., independent) are merged

**Fig. 8**. The DDBC Algorithm

## 2.2 Parallel Heterogeneous Clustering Algorithms

Heterogeneous clustering methods are utilized in cascade (i.e., end-result level) to cluster the dataset in parallel hybrid clustering. This cascading approach is mostly used to increase the quality of solutions created by a previous method (s). The hybrid clustering in the parallel mode is accomplished by either communicating with all the nodes or interacting with only facilitator nodes.

### 2.2.1 Parallel Hybrid PDDP and *k*-means (PDDP-KM) Algorithm

Figure 9 depicts the multiple stages in the parallel PDDP method [27]. The PKM improves the clustering solutions obtained by the parallel PDDP algorithm. End-result coordination between the concurrent PDDP and PKM underpins this hybrid combination. This collaboration technique primarily improves the solutions shaped by the PDDP algorithm and provides a strong starting point for the k-means method. The parallel PDDP-KM algorithm is shown in Fig. 10.

---

**Parallel PDDP Algorithm**
**Inputs:** The matrix M, the tree *height*, # nodes P
**Outputs:** *k* clusters
- The entire matrix M is the tree root
- Split the matrix M
*Stage1*:
   **For level *i* =1 to *height***
     **For cluster $S_i$ at level *i***
       If $S_i$ is a *singleton*, then process the next cluster; **<u>otherwise,</u>** $N_p$ computes the local mean vector $c^p_i$ of the cluster $S_i$, and the local leading eigenvector
         - Global centroid $c_i$ and leading eigenvector *u* are produced using MPI reduction call [32].
       **For each data point *x* in $S_i$**
         If (***u.x***) ≥ 0, allocate *x* to the left child of $S_i$, otherwise allocate *x* to the right child of $S_i$
*Stage2*: the desired set of *k* clusters is located at the leaf nodes.

---

**Fig. 9**. PDDP

---

**Algorithm: Parallel Hybrid PDDP - KM Clustering**
**Inputs**: The matrix M, the height of the tree, # nodes P, and # clusters *k*
**Outputs**: *k* clusters
Split M on the parallel P nodes
   *Stage1*: Use the PDDP to produce *k* centroids
   *Stage2*: Adopt the initial *k* centroids as input to the PKM
   *Stage3*: PKM produces global clusters *k*

---

**Fig. 10**. Parallel PDDP- KM Algorithm

## 3 Parallelization and Smart Solutions

In the era of the Internet of Things (IoT), various methods have been developed in Outlier detection [38][42], Recommendation Systems [43]-[48], Cyber Attacks Detection [49]-[52], Smart Systems [53]-[56], and forecasting methods [59]-[64]. Although centralized approaches in these fields have shown promises in improving the entire decision-making process using real-data or simulated data [65][67], scalability of these methods is a crucial step in real-time applications. Thus, applying split analysis using federated learning is mandatory to ensure proper implementations.

# 4 Conclusion and Future Directions

The scalability of traditional clustering methods is questioned when there is a need to cluster extensive and high-dimensional data. Parallel data clustering is an effective solution to cluster such enormous data. Several parallel clustering algorithms were presented in this paper; their clustering strategies were reviewed. New parallelization strategies have generated more efficient and challenging tasks. Different communication protocols (with different order of communication cost) have been established between nodes (or nodes and a facilitator node) to send local models or local data or local representatives and construct the final global model. Obviously, with the current trend in parallelizing computer architecture, it is more affordable and sensible to solve highly computationally intensive problems such as clustering in a parallel manner. Future directions would include the investigations of parallel performance measures that are well fit to the parallel environment compared to centralized measures [68].